\documentclass[aps,prmaterials,notitlepage,superscriptaddress,showkeys]{revtex4-2}

\usepackage{amsmath}
\usepackage{graphicx}	
\usepackage{rotating}

\usepackage{hyperref}	

\newcommand{\lef}{\left(}
\newcommand{\rig}{\right)}

\newcommand{\deriv}{\mathrm{d}}

\newcommand*{\rttensor}[1]{\bar{\bar{#1}}}

\newcommand{\Fermi}{\textsc{f}}

\newcommand{\veck}{\mathbf{k}}
\newcommand{\vecv}{\mathbf{v}}

\begin{document}

\title{First-principles-based screening method for resistivity scaling of anisotropic metals}

\author{Kristof Moors}
\email[Author to whom correspondence should be addressed. Email: ]{k.moors@fz-juelich.de}
\affiliation{Institute for Semiconductor Nanoelectronics, Peter Gr\"unberg Institute 9, Forschungszentrum J\"ulich, 52425 J\"ulich, Germany}

\author{Kiroubanand Sankaran}
\affiliation{Imec, 3001 Leuven, Belgium}
\author{Geoffrey Pourtois}
\affiliation{Imec, 3001 Leuven, Belgium}
\affiliation{University of Antwerp, Department of Chemistry, 2610 Wilrijk-Antwerpen, Belgium}
\author{Christoph Adelmann}
\email[Email: ]{christoph.adelmann@imec.be}
\affiliation{Imec, 3001 Leuven, Belgium}

\date{\today}

\begin{abstract}
The resistivity scaling of metals is a crucial limiting factor for further downscaling of interconnects in nanoelectronic devices that affects signal delay, heat production, and energy consumption. Here, we generalize a commonly considered figure of merit for selecting promising candidate metals with highly anisotropic Fermi surfaces in terms of their electronic transport properties at the nanoscale. For this, we introduce a finite-temperature transport tensor, based on band structures obtained from first principles. This transport tensor allows for a straightforward comparison between highly anisotropic metals in nanostructures with different lattice orientations and arbitrary transport directions. By evaluating the temperature dependence of the tensor components, we also assess the validity of a Fermi surface-based evaluation of the transport properties at zero temperature, rather than considering standard operating temperature conditions.
\end{abstract}


\maketitle

\section{\label{sec:introduction}Introduction}
The understanding of the electronic transport properties of metals under dimensional scaling has remained a relevant topic over many decades, as the continued downscaling of nanoelectronic devices requires ever narrower interconnects with low resistance at (and above) room temperature. A crucial property in this regard is the increase of the resistivity at reduced dimensions \cite{Kapur2002,Josell2009,Baklanov2014,Gall2020} due to surface or grain boundary scattering. This further aggravates the resistance increase of scaled interconnects, which is already unavoidable due to the shrinking cross-sectional area of wires or vias. This has led to an intensive effort to identify novel materials with lower resistivity at small dimensions, which also do not require barrier and liner layers for reliability \cite{Kapur2002,Josell2009,Baklanov2014,sankaran_2014,Gall2020,gall_resistivity_2020,adelmann_alternative_2014,adelmann_alternative_2018}. While new materials took and left the center stage over the years as material of choice for interconnects on the smallest length scales (e.g., from Al to Cu and Co) \cite{Kapur2002,edelstein_full_1997,auth_10nm_2017,dutta_sub-100_2018}, with others (e.g., Ru \cite{wen_atomic_2016,xunyuan_zhang_ruthenium_2016,Dutta2017}, or NiAl \cite{chen_nial_2018,soulie_thickness_2020}) in the research pipeline \cite{gall_materials_2021}, one of the major selection criteria for interconnect metals remains a low bulk resistivity that does not significantly increase as the interconnect dimensions are reduced.

For metals at near-room temperature conditions, the semiclassical Drude-Sommerfeld model remains to this day the canonical framework for understanding the electronic transport properties. Based on this framework, the resistivity scaling in metal thin films and wires was analyzed with ever increasing detail and refinement~\cite{Fuchs1938, Sondheimer1952,mayadas_electrical_1969, Mayadas1970,ke_resistivity_2009,moors_resistivity_2014,moors_modeling_2015,zhou_resistivity_2018,Gall2020} in terms of the various scattering processes that start to dominate the resistivity at nanoscale dimensions (e.g., grain boundary or surface roughness scattering). Without requiring a detailed treatment of scattering mechanisms, the product of resistivity $\rho$ and elastic mean free path (MFP) $\lambda$, the so-called $\rho\lambda$ product, already provides a useful figure of merit for gauging the sensitivity of the resistivity to scaled dimensions. Both quantities are bulk properties but their product provides a strong indicator for being highly conductive at reduced dimensions and is easily evaluated from first-principles-based band structure data. This product, evaluated at zero temperature, has therefore been commonly considered for screening and preselecting promising metals~\cite{Gall2016,Dutta2017,Gall2020, Sankaran2021}.

As more exotic metal materials and compounds with complicated band structures are being considered for nanoscale interconnect applications~\cite{Dutta2017,chen_nial_2018,chen_cual2_2019,chen_interdiffusion_2021,koike_intermetallic_2021,Sankaran2021,chen_potential_2021,soulie_aluminide_2021}, their resistivity (scaling) can be expected to depend strongly on the orientation of the interconnect line (if coarse-grained) with respect to the lattice orientation~\cite{choi_crystallographic_2013,choi_failure_2014,Zheng2017}, and on temperature.
For obtaining a meaningful figure of merit of anisotropic materials, the (distribution of) transport directions has to be specified when evaluating the $\rho\lambda$ product~\cite{Gall2016,DeClercq2018,Sankaran2021}. For novel exotic material that have yet to be materialized and for which the preferred growth direction or typical crystal texture is not yet known, this procedure is not ideal. A figure of merit that can be calculated \emph{a priori}, i.e., before specifying the transport direction or texture, would be preferred.

Here, we introduce a \emph{transport tensor} as a generalization of the $\rho\lambda$ product that naturally takes into account the symmetry group of the material and the anisotropy of the electronic band structure. The tensor can be calculated directly from first-principles-based band-structure data, without requiring a detailed treatment of the various scattering processes. To analyze the temperature dependence of the resulting figure of merit, we consider a finite-temperature evaluation of the transport tensor in contrast to previous works that consider the transport properties at zero temperature, only taking into account the charge-carrier states on the Fermi surface.

With these considerations, the transport tensor can serve as an indicator for low resistance at reduced dimensions and at operating temperature conditions around room temperature and above. It can easily be evaluated along arbitrary transport directions (with or without averaging over different directions) for many ``exotic'' metallic (compound) materials using standard \emph{ab initio} techniques. Furthermore, it can also provide a starting point for more detailed transport treatments of textured thin films or nanowires with confinement along particular directions with respect to the lattice orientation (see, e.g., Refs.~\cite{choi_crystallographic_2013} and \cite{Zheng2017}).

The article is structured as follows. In Sec.~\ref{sec:framework}, we present an overview of the theoretical framework. We apply this framework in Sec.~\ref{sec:results} to evaluate the transport tensor components for several example metals with different crystallographic symmetries. We discuss the results in Sec.~\ref{sec:discussion}, before concluding in Sec.~\ref{sec:conclusion}.

\section{\label{sec:framework}Theoretical framework}
\subsection{\label{subsec:transport}Semiclassical transport}
We start from the general semiclassical expression for the zero-frequency conductivity tensor $\rttensor{\sigma}$ of a bulk metal as a function of its electronic states with wave vector $\veck$, conduction band index $n$, energy $E^{(n)}(\veck)$, group velocity $\vecv^{(n)}(\veck) \equiv \nabla_\veck E^{(n)}(\veck)/\hbar$, considering the relaxation time approximation with relaxation time $\tau^{(n)}(\veck)$. The conductivity tensor can then be written as~\cite{Jacoboni2010}:

\begin{equation} \label{eq:conductivity_tensor}
	\begin{split}
	\rttensor{\sigma} &= -e^2 \sum_n \int_\textsc{bz} \frac{\deriv^3 \veck}{(2 \pi)^3} \vecv^{(n)}(\veck) \otimes \vecv^{(n)}(\veck) \tau^{(n)}(\veck) \left. \frac{\deriv f_\textsc{fd}(\epsilon)}{\deriv \epsilon} \right|_{\epsilon = E^{(n)}(\veck)} \\
	&\equiv \frac{e^2}{(2 \pi)^3} \sum_n \langle \vecv^{(n)}(\veck) \otimes \vecv^{(n)}(\veck) \tau^{(n)}(\veck) \rangle_\textsc{bz},
	\end{split}
\end{equation}

\noindent with $e$ the electron charge, $\hbar$ the reduced Planck's constant, and $f_\textsc{fd}(\epsilon) = (e^{(\epsilon - \mu)/(k_\textsc{b} T)} + 1)^{-1}$ the Fermi-Dirac distribution function with chemical potential $\mu$ and thermal energy $k_\textsc{b} T$. In the second line of Eq.~\eqref{eq:conductivity_tensor}, we introduce a notation that reflects that the conductivity tensor is a Fermi-Dirac-weighted average over a product of different velocity components and the relaxation time, summed over all bands with electronic states near the Fermi level. Note that the weight is not dimensionless but has a unit of inverse energy.
For conventional metals, the zero-temperature limit of this conductivity tensor is commonly considered. The integration over the three-dimensional $\veck$-space and Fermi-Dirac weight can then be replaced by an integration over the Fermi surface $S_\Fermi^{(n)}$ of each conduction band with index $n$:

\begin{equation} \label{eq:conductivity_tensor_T0}
	\begin{split}
	\rttensor{\sigma} &\overset{T = 0}{ = } \frac{e^2}{(2 \pi)^3 \hbar} \sum_n \int \deriv S_\Fermi^{(n)} \frac{\vecv^{(n)}(\veck) \otimes \vecv^{(n)}(\veck)}{|\vecv^{(n)}(\veck)|} \tau^{(n)}(\veck) \\
	&\equiv \frac{e^2}{(2 \pi)^3 \hbar} \sum_n \left\langle \frac{\vecv^{(n)}(\veck) \otimes \vecv^{(n)}(\veck)}{|\vecv^{(n)}(\veck)|} \tau^{(n)}(\veck) \right\rangle_{S_\Fermi^{(n)}}.
	\end{split}
\end{equation}

\noindent Note that we keep indices for spin degrees of freedom explicit in the notation. Hence, the band index $n$ should also include different spins. If all bands are spin degenerate, the explicit summation over the spin can be removed by multiplying the right-hand sides of Eqs.~\eqref{eq:conductivity_tensor} and \eqref{eq:conductivity_tensor_T0} with a factor of 2.

The general evaluation of the conductivity tensor requires an expression for the relaxation times $\tau^{(n)}(\veck)$, and the group velocities of all states with energies near the Fermi energy (within an energy window of $\sim k_\textsc{b} T$). The first requires information about electronic states and scattering mechanisms that they are subject to, while the second only depends on the electronic structure itself. The electronic structure can be described in different ways, with descriptions ranging from a completely isotropic effective mass approximation to a full-fledged atomistic band structure (\emph{e.g}., obtained from density functional theory calculations). For the examples considered here, we obtain the band structure from first-principles calculations (see below for details) to capture its anisotropy in full detail.

\subsection{\label{subsec:rholambda}Figure of merit $\rho\lambda$}
Starting from the general form of the conductivity tensor in Eq.~\eqref{eq:conductivity_tensor}, we can introduce approximations that allow for the calculation of the $\rho\lambda$ product, i.e., the figure of merit of interest, without any detailed knowledge of relaxation times \cite{Gall2016, Dutta2017,Gall2020, Sankaran2021}. A first approach is the assumption of a constant isotropic bulk MFP, $\vecv^{(n)}(\veck) \tau^{(n)}(\veck) = \lambda$. Note that this is implicitly an assumption for the relaxation time distribution $\tau^{(n)}(\veck)$. With this assumption, we can eliminate $\tau^{(n)}(\veck)$ from the conductivity tensor and write the following \emph{transport tensor}:

\begin{equation}
	\label{eq:rho_lambda}
	\rttensor{\lef \frac{1}{\rho \lambda} \rig} = \frac{e^2}{(2 \pi)^3} \sum_n \left\langle \frac{\vecv^{(n)}(\veck) \otimes \vecv^{(n)}(\veck)}{|\vecv^{(n)}(\veck)|} \right\rangle_\textsc{bz}.
\end{equation}

\noindent This tensor is a natural generalization of the (inverse) $\rho\lambda$ product that only depends on the group velocities of the electronic states.

Alternatively, one can consider the constant relaxation time approximation, $\tau^{(n)}(\veck) = \tau$, which naturally yields the following transport tensor:

\begin{equation}
	\label{eq:rho_tau}
	\rttensor{\lef \frac{1}{\rho \tau} \rig} = \frac{e^2}{(2 \pi)^3} \sum_n \langle \vecv^{(n)}(\veck) \otimes \vecv^{(n)}(\veck) \rangle_\textsc{bz}.
\end{equation}

In this case, it is $\rho\tau$ rather than $\rho\lambda$ that naturally generalizes to a tensor. To distill a directional figure of merit from this tensor, one can divide the tensor by the Fermi-Dirac weighted average velocity magnitude over all the bands, $v$, which can be obtained as follows:

\begin{equation}
	v \equiv \frac{\sum_n \langle |\vecv^{(n)}(\veck)| \rangle_\textsc{bz}}{\sum_n \langle 1 \rangle_\textsc{bz}}.
\end{equation}

We will refer to the resulting tensor as the $\rho v \tau$ tensor, as compared to the $\rho\lambda$ tensor.

Note that $v \tau$ represents an average MFP over all directions and does not take into consideration any particular transport direction. However, in certain cases, a more appropriate figure of merit can be constructed by considering a directional MFP. For example, the in-plane MFP should be small to reduce the impact of transverse grain boundaries on the resistivity, while the out-of-plane MFP should be small to reduce the impact of surface roughness. In Ref.~\cite{Sankaran2021}, for example, the in-plane-projected velocity is considered to obtain a figure of merit from the $\rho\tau$ product. Such a construction can only be considered after specifying the transport direction, however. Here, we opt for the MFP averaged over all directions to retain the tensor form of the figure of merit, which only needs to be calculated once (without requiring the transport direction to be specified) and can afterwards be evaluated along arbitrary transport directions. We expect that such a figure of merit is already sufficient for a preselection of the most promising highly anisotropic materials.

The $\rho\lambda$ and $\rho v \tau$ tensors retain the tensor structure of the conductivity and thereby capture the symmetry of the metal and its electronic band structure. To evaluate the sensitivity of the metallic resistivity to scaled dimensions for transport along the main crystallographic directions, the diagonal tensor components are sufficient. For the evaluation along arbitrary directions or averaging over different directions (e.g., in the case of columnar grains with arbitrary in-plane rotation of the lattice structure), the off-diagonal components (which may be positive or negative) become relevant as well. For example, the $\rho\lambda$ tensor transforms as $(1/\rho\lambda)_{uu} = \sum_{\alpha,\beta} (\partial \alpha/\partial u) (\partial \beta/\partial u) (1/\rho\lambda)_{\alpha\beta}$ along an arbitrary direction $u$, with $\alpha,\beta \in \{x, y, z\}$  (analogously for the figure of merit $1/\rho v \tau$).

We will evaluate and compare both variants of the transport tensor, obtained under the assumption of constant $\lambda$ or constant $\tau$, for different example metals below. Details on the numerical integration scheme for the transport tensors can be found in Appendix~\ref{sec:numerical_integration}.

\subsection{\label{subsec:band_structure}First-principles calculations of band structures}
The computation of the electronic band structures were performed using density functional theory simulations, as implemented in the \textsc{quantum espresso} package~\cite{Giannozzi2009} together with the valence electrons represented by Garrity-Bennett-Rabe-Vanderbilt pseudopotentials~\cite{Garrity2014} with a kinetic cutoff between 60 and 80 Ry (depending on the elemental composition) including a Methfessel-Paxton smearing function with a broadening of 13.6 meV.
The valence electrons are evaluated within the Perdew-Burke-Ernzerhof generalized gradient approximation~\cite{Perdew1996} together with the first Brillouin zone sampled using a regular Monkhorst-Pack scheme~\cite{Monkhorst1976} with a $k$-point density ranging from 25$\times$25$\times$25 to 61$\times$61$\times$61, depending on the material under consideration.
This ensures a convergence of the total energy within 10$^{–12}$ eV and allows for a smooth interpolation of the band structure.
The atomic relaxation of the geometric configurations is continued until all residual atomic forces are smaller than 10$^{-4}$ eV/Å.

We consider a number of example metals for the evaluation of the transport tensor, representing all the different symmetry families (see subsection below). Some of their properties (Fermi energy, lattice constant, number of bands near the Fermi level) are listed in Table~\ref{table:EF}. The group of example metals covers both reference metals (Cu), as well as metals that were proposed as alternatives for interconnect metallization (Ru \cite{wen_atomic_2016,xunyuan_zhang_ruthenium_2016,Dutta2017}, V$_2$AlC) in addition to novel binary intermetallics. The results below show that MoPt is of potential interest, with a figure of merit that can outperform Cu for transport along certain directions.

\subsection{\label{subsec:symmetry_groups}Symmetry families}
The detailed form of the conductivity tensors can be related to the different crystal systems of the metals and their symmetries \cite{Powell2010}, assuming that these symmetries are not broken by the relaxation time profile $\tau^{(n)}(\veck)$. In general, the conductivity tensor has the following form:

\begin{equation} \label{eq:general_form}
	\rttensor{\sigma} = \begin{pmatrix}
		\sigma_{xx} & \sigma_{xy} & \sigma_{xz} \\
		\sigma_{yx} & \sigma_{yy} & \sigma_{yz} \\
		\sigma_{zx} & \sigma_{zy} & \sigma_{zz}
	\end{pmatrix}.
\end{equation}

The tensor is symmetric by construction ($\sigma_{ij} = \sigma_{ji}$), and its components are further restricted by the symmetry family of the material lattice. Table~\ref{table:symm_groups} represents an overview of the different symmetry families and the corresponding general forms of the conductivity tensors with the number of independent components, as well as some example metals for each symmetry family.

\section{\label{sec:results}Results}
\subsection{\label{subsec:group_velocity}Group velocity}
\begin{figure}
	\centering
	\includegraphics[width=8cm]{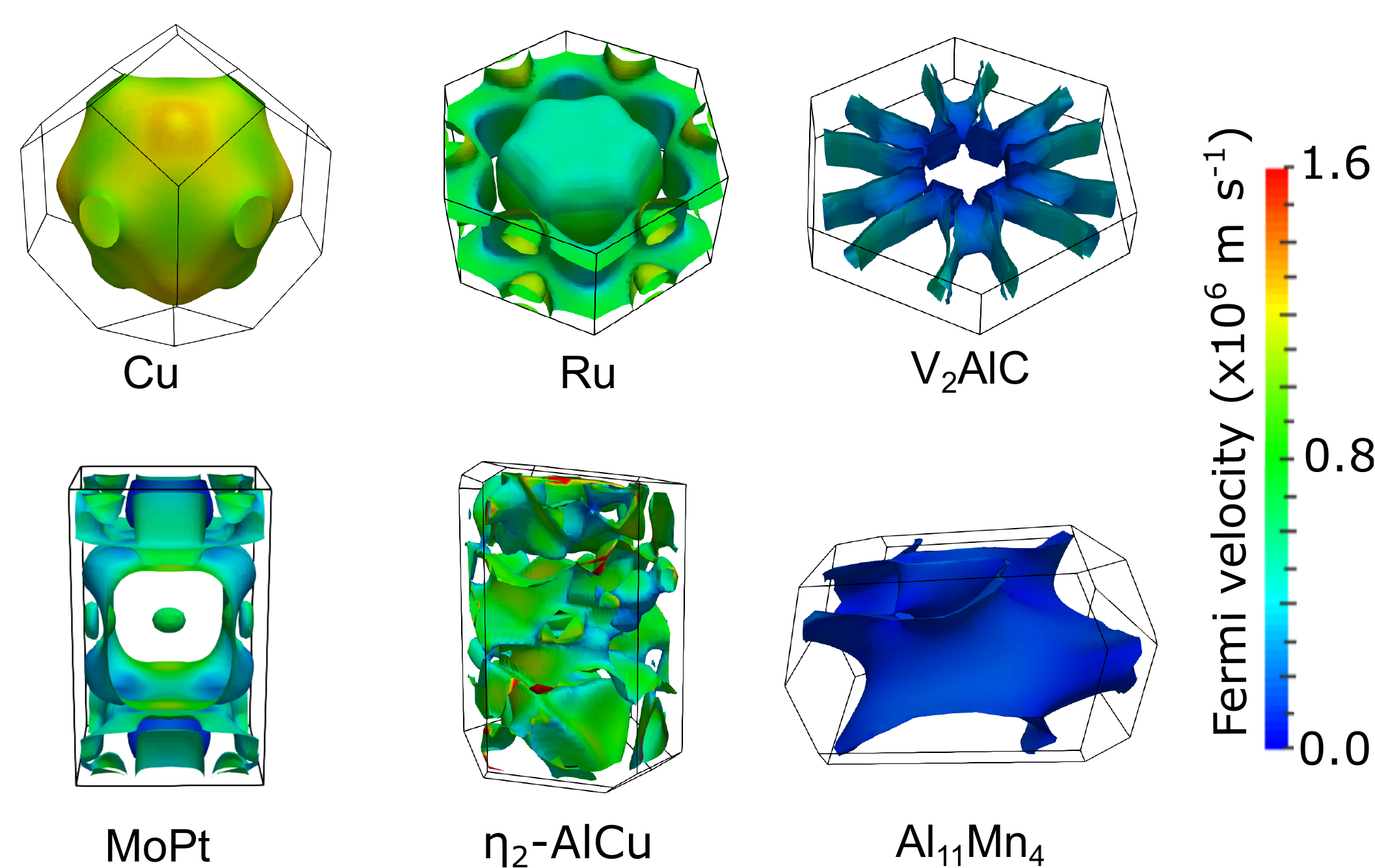}
	\caption{
		Fermi surfaces of the example metals with different symmetry families listed in Table~\ref{table:symm_groups}. The magnitude of the Fermi velocity is indicated by color on the Fermi surface.
	}
	\label{fig:1}
\end{figure}

The evaluation of the transport tensors in Eqs.~\eqref{eq:rho_lambda} and \eqref{eq:rho_tau} requires the group velocities of all electronic states near the Fermi level. Here, we present the group velocities for the example metals that are listed in Table~\ref{table:symm_groups}, which are obtained from first-principles calculations (see Sec.~\ref{subsec:band_structure}). In Fig.~\ref{fig:1}, the magnitude of the group velocity is indicated by color for all electronic states on the Fermi surface. The distributions of the group velocity magnitude are shown in Fig.~\ref{fig:2}, as well as the distributions of the different components, of which the (an)isotropy can be clearly seen. A detailed overview of the velocity distributions of the different conduction bands is presented in Appendix~\ref{sec:velocity_distributions}.

\subsection{\label{subsec:transport_tensors}Transport tensors}

Having obtained the group velocities of all the electronic states near the Fermi level, the $1/\rho \lambda$ and $1/\rho v \tau$ transport tensors can now be numerically evaluated in a straightforward manner using Eqs.~\eqref{eq:rho_lambda} and \eqref{eq:rho_tau}. In this way, we obtain a generalization of the $\rho\lambda$ figure of merit along different directions and under assumptions of a constant bulk relaxation time or MFP.
Figure~\ref{fig:3} represents the tensor components for the example metals in Table~\ref{table:EF}, relative to the tensor component with the largest magnitude. Here, Fermi-Dirac statistics at $300\,\textnormal{K}$ were considered in all calculations. The presented fractions are rounded to the nearest multiple of $0.025$, in keeping with the numerical accuracy. We note that, within this accuracy, the obtained tensor has the form that is expected from symmetry considerations of the material under consideration.

Figure~\ref{fig:4} illustrates the results for the figure of merit along the $x$, $y$, and $z$ directions, evaluated at $T = 50\,\textnormal{K}$ and $T = 300\,\textnormal{K}$ for comparison. The results indicate that, in general, the $1/\rho\lambda$ and $1/\rho v \tau$ tensor components are in good qualitative agreement. Hence, they can both be considered for a preselection of promising materials. Nonetheless, noticeable differences appear when comparing the tensors for the highly anisotropic example materials. This is expected, however, as the assumptions of constant relaxation time and constant MFP become more distinct as the anisotropy increases.

\subsection{\label{subsec:temperature_dependence}{Temperature dependence}}

\begin{figure*}
	\centering
	\includegraphics[width=16cm]{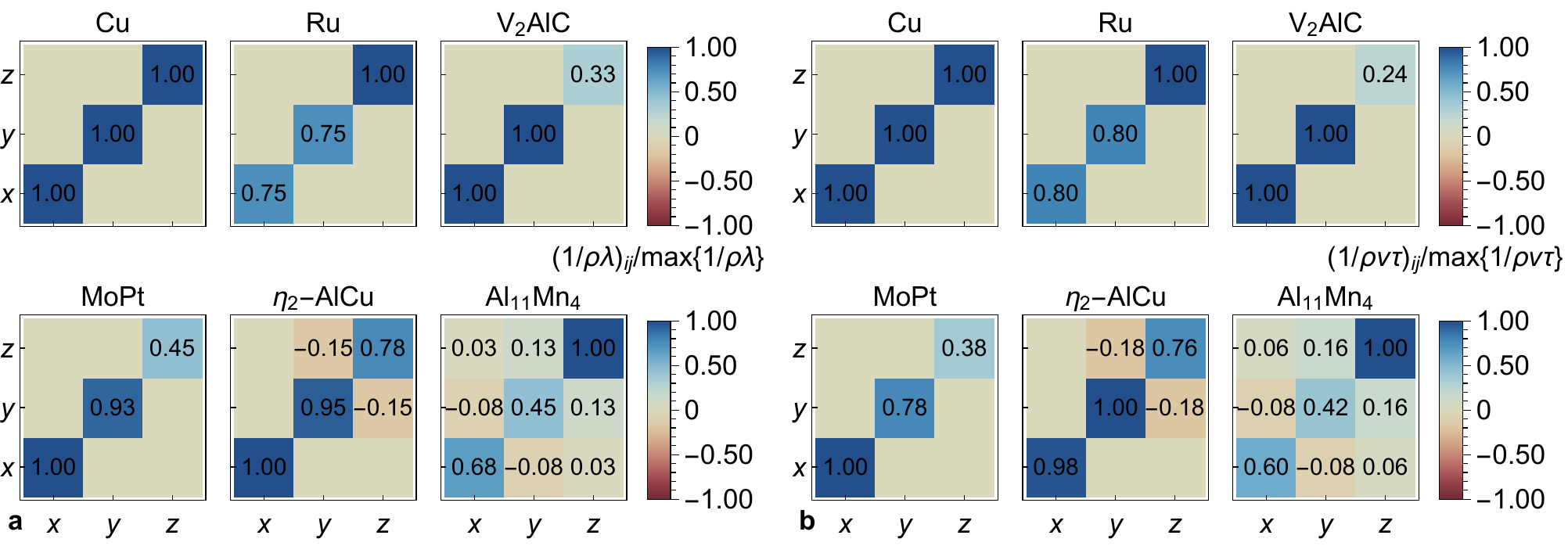}
	\caption{
		The values of the (a) $1/\rho \lambda$ and (b) $1/\rho v \tau$ tensor components for the example metals of Table~\ref{table:EF}, evaluated at $300\,\textnormal{K}$, relative to the tensor component with the largest magnitude.
	}
	\label{fig:3}
\end{figure*}

\begin{figure*}
	\centering
	\includegraphics[width=16cm]{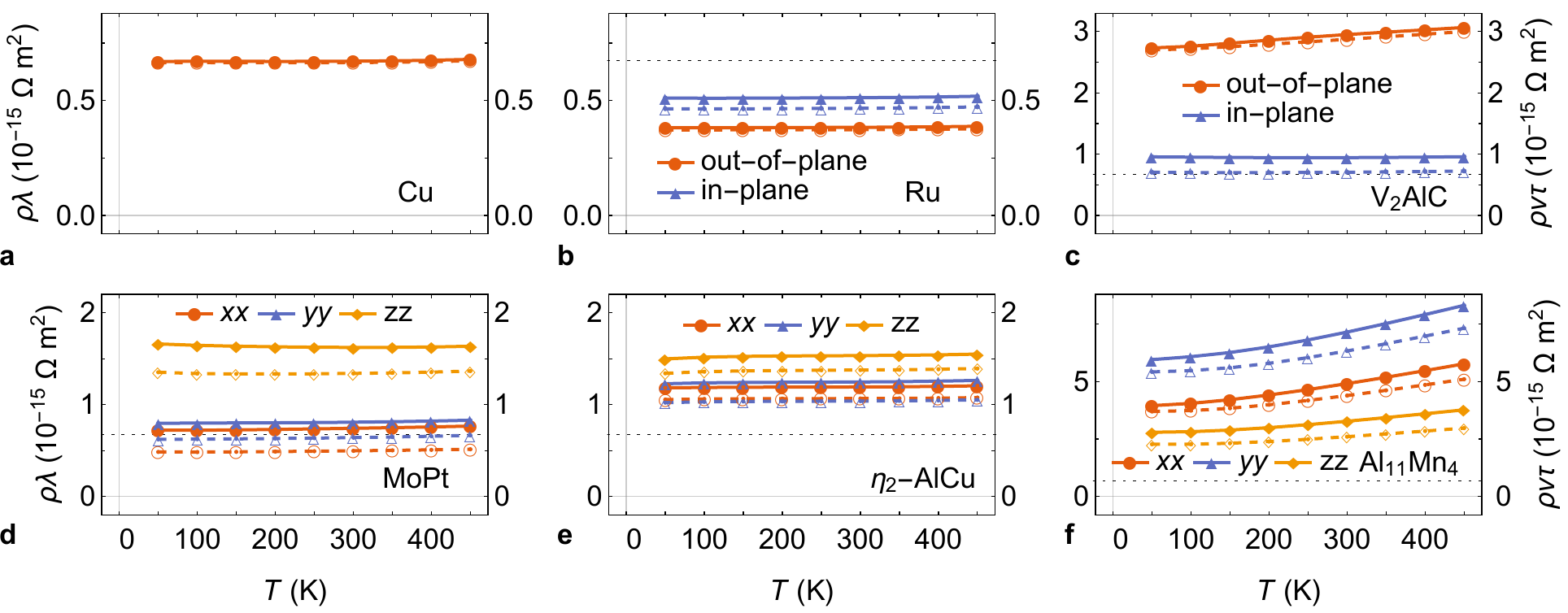}
	\caption{
		The (inverse) values of the diagonal $1/\rho \lambda$ (solid lines, scale on the left) and $1/\rho v \tau$ (dashed lines, scale on the right) tensor components for the example metals listed in Table~\ref{table:symm_groups} as a function of temperature.
		The black dotted line indicates the result for Cu as a reference.
	}
	\label{fig:5}
\end{figure*}

\begin{figure}
	\centering
	\includegraphics[width=8cm]{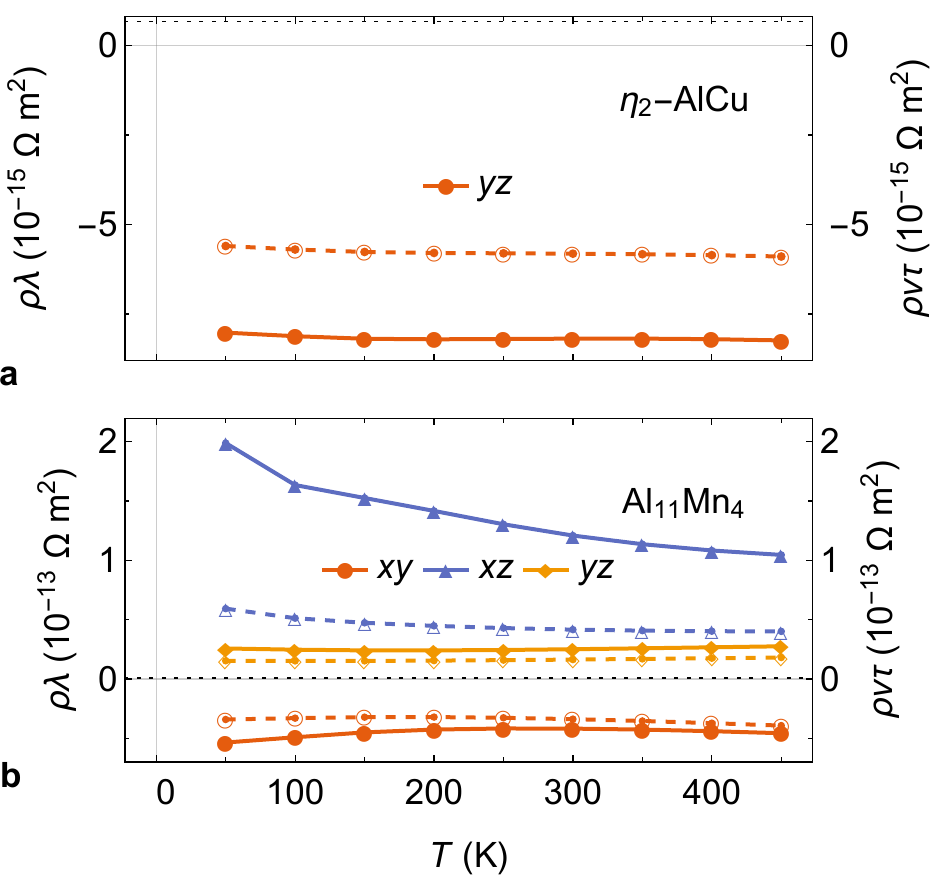}
	\caption{
		(a),(b) The (inverse) values of the off-diagonal components of the $1/\rho \lambda$ (solid lines, scale on the left) and $1/\rho v \tau$ (dashed lines, scale on the right) for (a) $\eta_2$-AlCu and (b)  Al$_{11}$Mn$_4$ as a function of temperature.
		The black dotted line indicates the result for Cu as a reference.
	}
	\label{fig:6}
\end{figure}

Finally, we take a closer look at the temperature dependence of the tensor components. Results of all $\rho\lambda$ and $\rho v \tau$ tensor components at $T = 300\,\textnormal{K}$ are shown in Fig.~\ref{fig:3}. The temperature dependence of the diagonal tensor components of the different example materials are presented over a range between $T = 50\,\textnormal{K}$ and $T = 450\,\textnormal{K}$ in Fig.~\ref{fig:5}, and the off-diagonal components of $\eta_2$-AlCu and Al$_{11}$Mn$_4$ in Fig.~\ref{fig:6}. We can see that there is no significant temperature dependence of the figure of merit for most example materials, as can be expected for metals with $k_\textsc{b} T \ll E_\textsc{f}$. Notable exceptions, however, are the values for all components of Al$_{11}$Mn$_4$ (Figs.~\ref{fig:5}f and \ref{fig:6}b), and the values along the out-of-plane and $z$ directions for V$_2$AlC (Fig.~\ref{fig:5}c) and MoPt (Fig.~\ref{fig:5}d), respectively.

A temperature dependence arises when the band structure morphology (in particular, the velocity distribution) changes significantly in an energy window $\sim k_\textsc{b} T$ around the Fermi energy. This is most likely to occur when the Fermi velocity is small such that a linearization of the spectrum with constant velocity, $E^{(n)}(\veck + \delta \veck) \approx E_\textsc{f} + \hbar \, \delta \veck \cdot \vecv^{(n)}(\veck)$ [with $E^{(n)}(\veck) = E_\textsc{f}$], only provides a good approximation in a small energy window around $E_\textsc{f}$. Indeed, the example materials with the smallest mean velocity magnitude are precisely those that display a noticeable temperature dependence.

To recognize and predict a significant temperature dependence for the figure of merit on a more quantitative level, we can look at the total Brillouin zone volume that is occupied by the bands as a function of temperature, $\mathcal{V}^{(n)}$, as weighted by Fermi-Dirac statistics:

\begin{equation}
	\mathcal{V}^{(n)} = \int_\textsc{bz} \deriv^3 \veck \left. \frac{\deriv f_\textsc{fd}(\epsilon)}{\deriv \epsilon} \right|_{\epsilon = E^{(n)}(\veck)}.
\end{equation}

\begin{figure*}
	\centering
	\includegraphics[width=16cm]{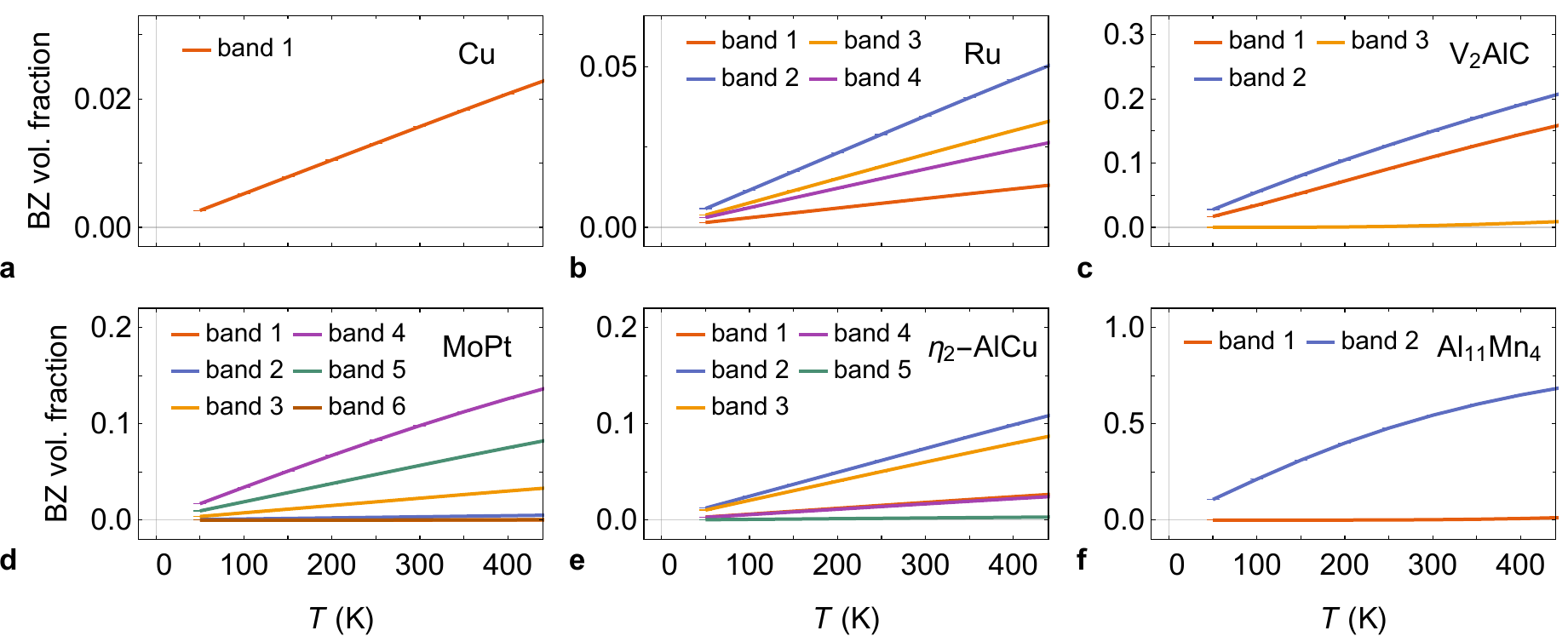}
	\caption{
		The fraction of the total Brillouin zone volume that is occupied by each band, weighted by Fermi-Dirac statistics, as a function of temperature, for the example metals listed in Table~\ref{table:symm_groups} .
	}
	\label{fig:7}
\end{figure*}

\noindent When $k_\textsc{b} T \ll E_\textsc{f}$ and the energy spectrum is well approximated by a linearized spectrum, the (reciprocal) volume can be obtained from the Fermi surface $S_\textsc{f}^{(n)}$ as follows: $\mathcal{V}^{(n)} \approx 4 k_\textsc{b} T S_\textsc{f}^{(n)} / (\hbar v)$, with $v$ the average velocity evaluated at zero temperature (the average over the Fermi surface). Hence, the occupied volume should be linear as a function of temperature. In Fig.~\ref{fig:7}, we show the fraction of the total Brillouin zone volume that is occupied for the different bands of the example metals with electronic states near the Fermi level. It can be clearly seen that most bands display a linear relation between the occupied volume and temperature. For certain bands of some materials, however, there is a noticeable nonlinearity. It is most pronounced for Al$_{11}$Mn$_4$ (Fig.~\ref{fig:7}f), and a weak nonlinearity can also be identified for the bands of V$_2$AlC (Fig.~\ref{fig:7}c) and MoPt (Fig.~\ref{fig:7}d) with the largest volume fraction. For all example metals with a noticeable nonlinearity in the $\mathcal{V}^{(n)}(T)$ relation, the tensor components also display a noticeable temperature dependence. As expected, these properties are strongly correlated.

\section{\label{sec:discussion}Discussion}

\begin{figure}
	\centering
	\includegraphics[width=8cm]{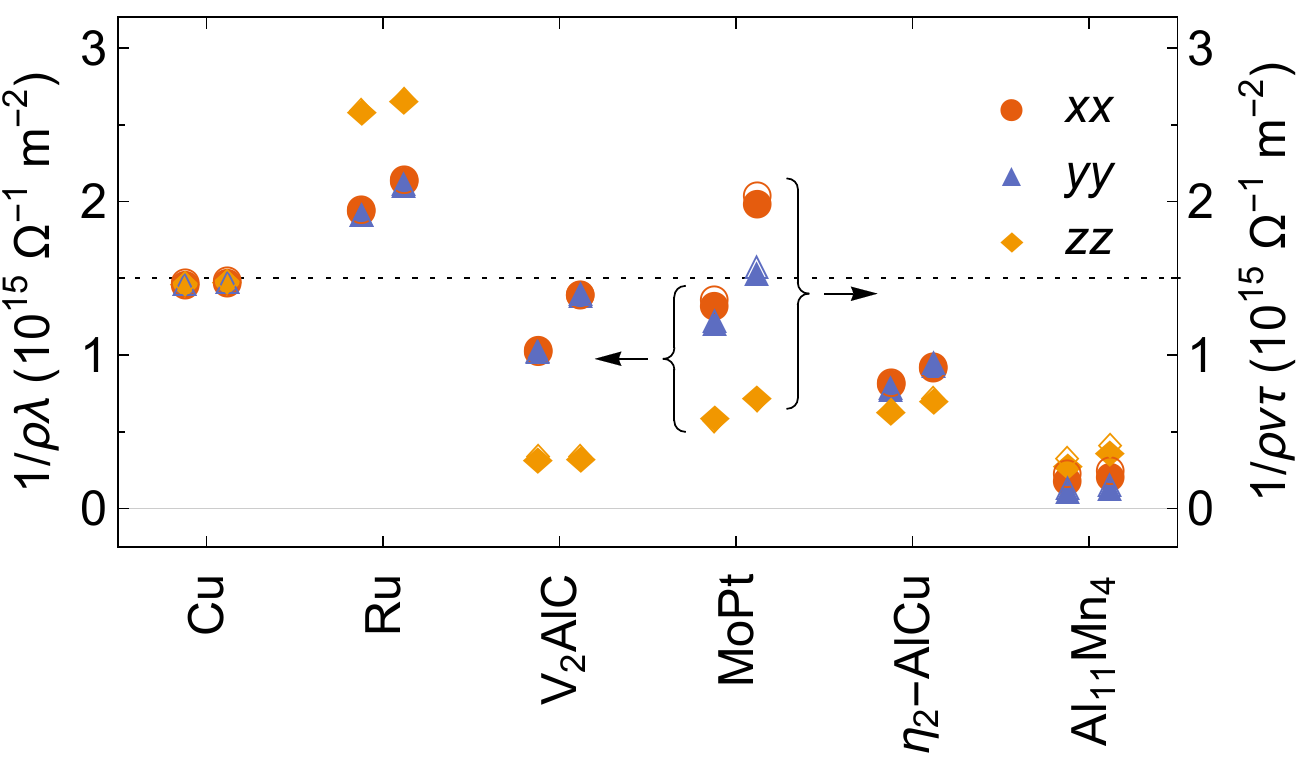}
	\caption{
		The value of $1/\rho \lambda$ (data points on the left) and $1/\rho v \tau$ (data points on the right), evaluated along the $x$, $y$, and $z$ directions, for the example metals of Table~\ref{table:EF}. The results are obtained from transport tensors with Fermi-Dirac statistics at $T = 50\,\textnormal{K}$ (hollow symbols) and $T = 300\,\textnormal{K}$ (filled symbols), which coincide in most cases.
		The black dotted line indicates the result for Cu as a reference.
	}
	\label{fig:4}
\end{figure}

As Cu has been a longstanding standard for nanoscaled interconnect applications, the (isotropic) $\rho\lambda$ value of Cu can be considered as a standard reference for material screening purposes of alternative metals. Based on the results in Fig.~\ref{fig:4}, it seems that three examples that are presented here have potential to outperform Cu at small dimensions: Ru (in all directions), V$_2$AlC (along the in-plane directions), and MoPt (along $x$ and $y$). 

\begin{figure}
	\centering
	\includegraphics[width=8cm]{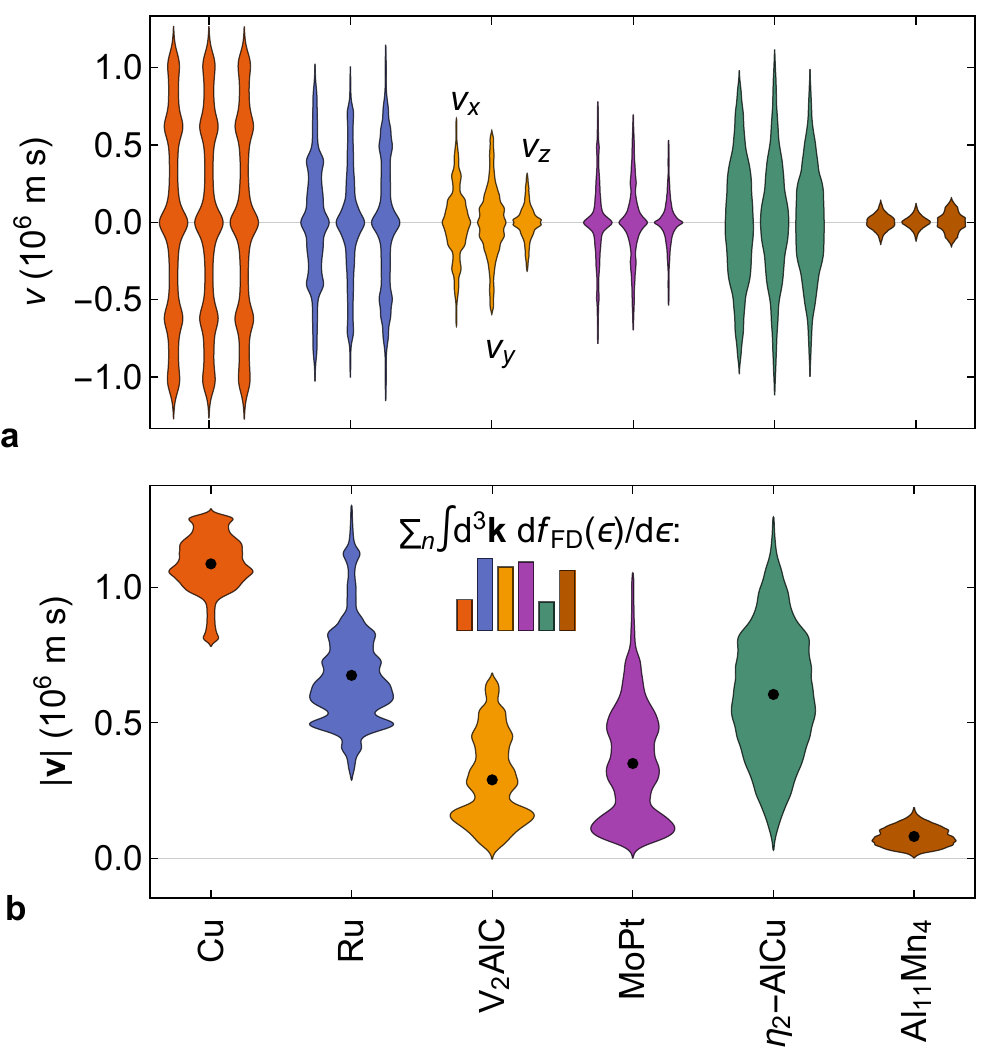}
	\caption{
		(a) The Fermi-Dirac-weighted distribution of group velocities along the $x$, $y$, and $z$ directions is presented for the example metals of Table~\ref{table:EF}.
		(b) The Fermi-Dirac-weighted distribution of group velocity magnitudes $|\vecv|$. The mean value is indicated by a black dot and the Fermi-Dirac-weighted volume of the charge carriers in reciprocal space is represented for the different metals by a bar chart in an inset.
		For the Fermi-Dirac statistics, we considered $T = 300\,\textnormal{K}$ and the Fermi energy listed in Table~\ref{table:EF}.
	}
	\label{fig:2}
\end{figure}

Interestingly, none of the promising alternative candidates have a Fermi velocity distribution that comes near the distribution of Cu, which has a very large mean value and a relatively low spread. As a matter of fact, large Fermi velocities contribute to long MFPs and are therefore simultaneously (and conflictingly) reducing (via a lower resistivity $\rho$) and increasing (via a longer MFP $\lambda$) the $\rho\lambda$ product.
However, the group velocity is not the only relevant quantity for a promising $\rho\lambda$ value. A second key property is a high carrier density, as can be seen in Eqs.~\eqref{eq:rho_lambda} and \eqref{eq:rho_tau}. A high carrier density is reflected by a large Fermi-Dirac-weighted volume in reciprocal space, which is depicted  in the inset in Fig.~\ref{fig:2}b. When these two aspects are combined, it becomes clear that the promising alternative metal candidates compensate an overall smaller (average) Fermi velocity (compared to Cu) with a significantly higher charge-carrier density. In other words, promising alternative metals combine a high carrier concentration with a carrier mobility that is low enough to lead to a short MFP (compared to Cu), but is not so low to be detrimental to the resistivity. The $\rho\lambda$ value is further lowered along certain directions by an anisotropic group (Fermi) velocity distribution. 

While the validity of the constant MFP approximation in Eq.~\eqref{eq:rho_lambda} or the constant relaxation time approximation in Eq.~\eqref{eq:rho_tau} in resistivity scaling models for thin films and nanowires has not been proven and is in general questionable, it appears that the two transport tensors based on the assumption of constant $\lambda$ or constant $\tau$, respectively, are generally in good qualitative agreement, even for highly anisotropic materials. In particular, the comparison with the isotropic Cu values does not appear to depend on the approximation, a conclusion that was also reported for a large number of ternary MAX materials \cite{Sankaran2021}. This suggests that \emph{ab initio} screening methods based on $\rho\lambda$ or $\rho v \tau$ products are to some extent ``robust'' since they do not depend on the exact approximation made. The results further suggest that the details of the relaxation time distribution do not particularly matter for material screening purposes and both transport tensors are equally suited in this regard.

Finally, we would like to comment on the limitations of the presented screening methodology. A first limitation is that all details about the dominant scattering mechanism upon introducing confinement are neglected. Currently, no Mayadas-Shatzkes-equivalent transport model for thin-film transport with grain boundary and boundary surface scattering is available that incorporates all the details of the band structure or Fermi surface. Such a model would be a key step forward to assess the resistivity screening of alternative metals with complex and anisotropic band structures. The availability of such a model remains a major milestone for a refined assessment of the resistivity of metals at small dimensions.
Some recent works go in this direction. One approach is to describe the band-structure anisotropy with an anisotropic effective mass tensor, for which a generalization of the Mayadas-Shatzkes model can be obtained analytically, as presented in Ref.~\cite{DeClercq2018}. Another approach is presented in Ref.~\cite{Zheng2017}, which generalizes the Fuchs-Sondheimer model for the conductivity in thin films limited by boundary surface scattering to arbitrary Fermi surfaces. This model can be applied to single-crystal films without any grains and yields the following expression for the conductivity when assuming a constant MFP $\lambda$ and thin-film confinement along $z$ with fully diffuse surface scattering (writing the zero-temperature expression here for convenience):

\begin{equation}
	\sigma = \frac{e^2}{\hbar} \lambda \sum_n \int \frac{\deriv S_\Fermi^{(n)}}{(2 \pi)^3} \frac{v_x^{(n)}(\veck)^2}{|\vecv^{(n)}(\veck)|^2} \left\{ 1 + \left[ \exp\left(-\frac{d|\vecv^{(n)}(\veck)|}{\lambda v_z^{(n)}(\veck)}\right) - 1 \right] \frac{\lambda v_z^{(n)}(\veck)}{d|\vecv^{(n)}(\veck)|} \right\},
\end{equation}

\noindent with $d$ the film thickness. This conductivity can be generalized to a \emph{thin-film transport tensor} with respect to two independent in-plane directions (denoted by subscripts $\alpha$ and $\beta$ for the in-plane velocity components below):

\begin{equation}
	\sigma_{\alpha\beta} = \frac{e^2}{\hbar} \lambda \sum_n \int \frac{\deriv S_\Fermi^{(n)}}{(2 \pi)^3} \frac{v_\alpha^{(n)}(\veck) v_\beta^{(n)}(\veck)}{|\vecv^{(n)}(\veck)|^2} \left\{ 1 + \left[ \exp\left(-\frac{d|\vecv^{(n)}(\veck)|}{\lambda v_\perp^{(n)}(\veck)}\right) - 1 \right] \frac{\lambda v_\perp^{(n)}(\veck)}{d|\vecv^{(n)}(\veck)|} \right\},
\end{equation}

\noindent with $v_\perp^{(n)}(\veck)$ the velocity component along the normal of the thin film. The symmetries of this tensor are determined by the crystal family of the material while also taking into account the in-plane versus out-of-plane symmetry breaking due to the thin-film confinement. Note that the constant mean free path $\lambda$ cannot be pulled out of the integral in this case, which prevents us from constructing a $1/\rho\lambda$ tensor.
A second limitation of the current metal screening methodology is that a metal becomes only promising when it combines a small $\rho\lambda$ figure of merit with a low bulk resistivity $\rho$. For many of the more ``exotic'' materials, bulk resistivities are unknown or reports may be conflicting. At present, for more refined screening, one can then consider a first-principles-based calculations of the electron-phonon scattering rate and plug this into Eq.~\eqref{eq:conductivity_tensor} to evaluate the bulk conductivity tensor directly, in combination with the transport tensors in Eqs.~\eqref{eq:rho_lambda} and \eqref{eq:rho_tau}.
A third limitation is that our transport formalism neglects any effects due to quantum confinement. However, these effects only become relevant for metals with very high carrier densities and extremely reduced dimensions (few-nm regime) in all transverse directions, as considered in Refs.~\cite{moors_resistivity_2014} and \cite{Lanzillo2017}, for example, with semiclassical and (atomistic) quantum transport approaches, respectively.
We would like to stress, however, that all the above-mentioned limitations can be investigated with a more specialized theoretical framework after preselecting the most promising materials with the generally applicable screening method presented here.

\section{\label{sec:conclusion}Conclusion}

In conclusion, we introduced transport-related tensors that generalize the previously introduced $\rho\lambda$ product as a figure of merit for the screening of metallic systems along arbitrary transport directions for advanced nanoscale conductor applications at standard operating temperature conditions around room temperature \cite{Gall2016,Dutta2017,Gall2020,Sankaran2021}. 
These tensors can be evaluated based on \emph{ab initio} band-structure data near the Fermi level and on generic assumptions for the dominant scattering processes of the system under consideration.
A comparison between different assumptions for the relaxation time and analysis of the temperature dependence shows that the tensorial figure of merit is robust and suitable for materials with a highly anisotropic electronic band structure.
Together with additional proxies such as the cohesive energy for the metal reliability as well as the bulk resistivity, this approach provides a suitable and generic tool for screening and benchmarking novel exotic (non-cubic) metals for scalable device applications that require excellent electronic transport properties subsisting at the nanoscale.

\begin{acknowledgments}
	This work has been supported by imec's industrial affiliate program on Nano-Interconnects.
	K.M.\ acknowledges the financial support by the Bavarian Ministry of Economic Affairs, Regional Development and Energy within Bavaria’s High-Tech Agenda Project “Bausteine für das Quantencomputing auf Basis topologischer Materialien mit experimentellen und theoretischen Ansätzen” (Grant No.\ 07 02/686 58/1/21 1/22 2/23).
\end{acknowledgments}

\bibliography{2021_Moors_transport_tensors}

\clearpage

\begin{sidewaystable}[h]
	\caption{\label{table:EF}Properties of selected example metals (PV stands short for primitive vector in reciprocal space).}
	\begin{ruledtabular}
		\begin{tabular}{l l l c c c c c c}
			Metal & Crystal symmetry & (symmetry group) & $E_\textsc{f}$ (eV) & $a_\text{lat}$ (\AA$^{-1}$) & \# bands near $E_\textsc{f}$ & PV I ($2\pi/a_\text{lat}$) & PV II ($2\pi/a_\text{lat}$) & PV III ($2\pi/a_\text{lat}$) \\
			\colrule
			Cu & cubic & (Fm$\bar{3}$m - 225) & 14.66 & 3.61 & 1 & $(-1,-1,1)$ & $(1,1,1)$ & $(-1,1,-1)$ \\
			Ru & hexagonal & (P6$_3$/mmc - 194) & 16.80 & 2.71 & 4 & $(1,0.58,0)$ & $(0,1.15,0)$ & $(0,0,0.63)$ \\
			V$_2$AlC & hexagonal & (P6$_3$/mmc - 194) & 13.75 & 2.90 & 3 & $(1,0.58,0)$ & $(0,1.15,0)$ & $(0,0,0.22)$ \\
			MoPt & orthorhombic & (Pmma - 51) & 22.26 & 2.75 & 6 & $(1,0,0)$ & $(0,0.61,0)$ & $(0,0,0.55)$ \\
			$\eta_2$-AlCu & monoclinic & (C2/m - 12) & 11.27 & 12.08 & 5 & $(1,-0.11,-0.07)$ & $(0,0.68,-0.25)$ & $(0,0,0.72)$ \\
			Al$_{11}$Mn$_4$ & triclinic & (P$\bar{1}$ - 2) & 10.55 & 5.03 & 2 & $(1,0.18,0.05)$ & $(0,1.01,0.28)$ & $(0,0,0.59)$ \\
		\end{tabular}
	\end{ruledtabular}
\end{sidewaystable}

\clearpage

\begin{sidewaystable}
	\caption{\label{table:symm_groups}Crystal families and their conductivity tensor.}
	\begin{ruledtabular}
		\begin{tabular}{l | c c c c c}
			Crystal family: & Triclinic & Monoclinic & Orthorhombic & (Tetra-Tri-Hexa)gonal & Cubic \\
			\colrule
			Conductivity tensor: & $\begin{pmatrix}
				\sigma_{xx} & \sigma_{xy} & \sigma_{xz} \\
				\sigma_{xy} & \sigma_{yy} & \sigma_{yz} \\
				\sigma_{xz} & \sigma_{yz} & \sigma_{zz}
			\end{pmatrix}$ & $\begin{pmatrix}
				\sigma_{xx} & 0 & \sigma_{xz} \\
				0 & \sigma_{yy} & 0 \\
				\sigma_{xz} & 0 & \sigma_{zz}
			\end{pmatrix}$ & $\begin{pmatrix}
				\sigma_{xx} & 0 & 0 \\
				0 & \sigma_{yy} & 0 \\
				0 & 0 & \sigma_{zz}
			\end{pmatrix}$ & $\begin{pmatrix}
				\sigma_\parallel & 0 & 0 \\
				0 & \sigma_\parallel & 0 \\
				0 & 0 & \sigma_\perp
			\end{pmatrix}$ & $\begin{pmatrix}
				\sigma & 0 & 0 \\
				0 & \sigma & 0 \\
				0 & 0 & \sigma
			\end{pmatrix}$ \\
			Tensor components (indep.): & 6 & 4 & 3 & 2 & 1 \\
			Examples (see Fig.~\ref{fig:1}): & Al${}_{11}$Mn${}_4$ & $\eta_2$-AlCu & MoPt & Ru (hcp), V${}_2$AlC & Cu (fcc)
		\end{tabular}
	\end{ruledtabular}
\end{sidewaystable}

\clearpage

\clearpage

\appendix
\renewcommand\thefigure{A\arabic{figure}}
\setcounter{figure}{0}

\section{\label{sec:velocity_distributions}Velocity distributions}

\begin{figure}
	\centering
	\includegraphics[width=16cm]{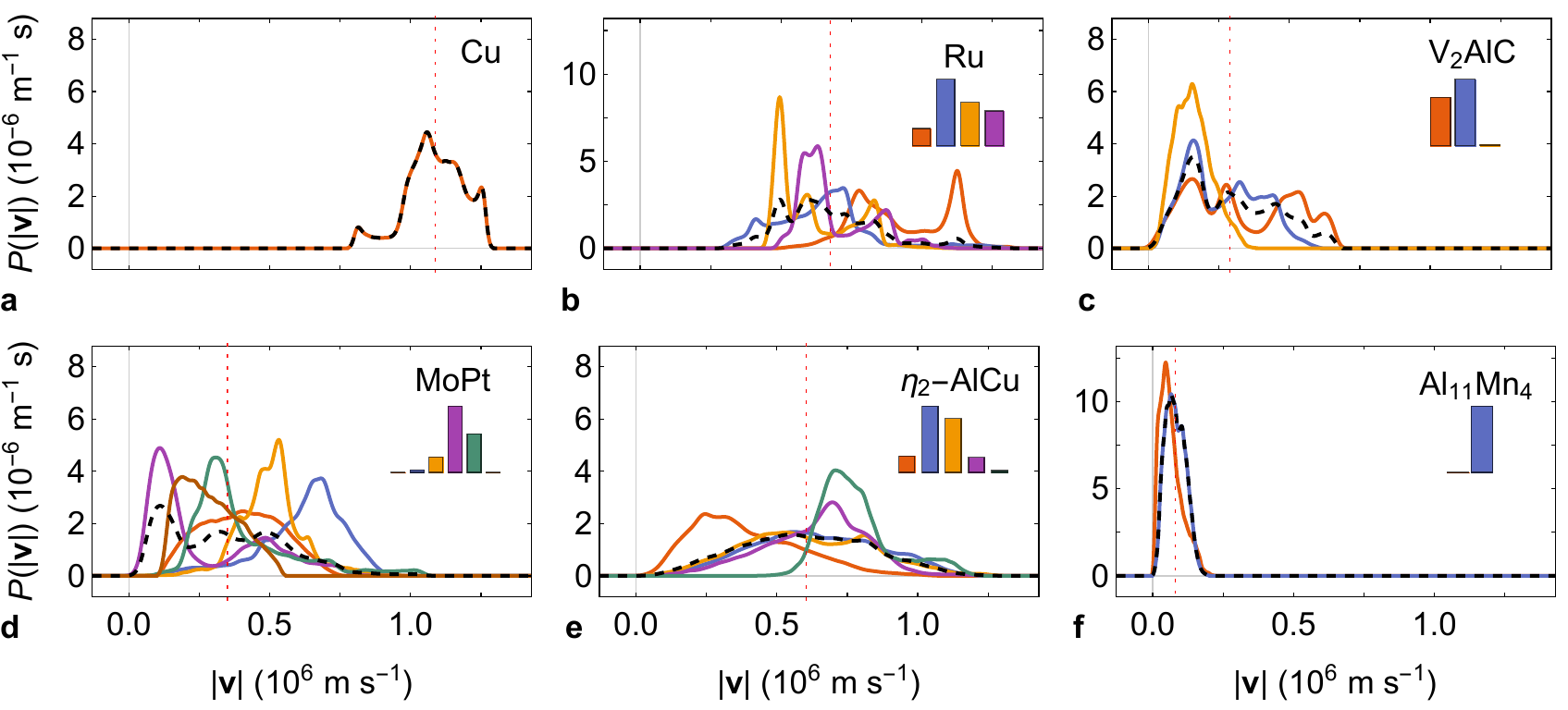}
	\caption{
		The distribution $P(|\vecv|)$ of group velocity magnitudes is presented for the different bands with states near the Fermi energy (indicated by different colors) of the example metals listed in Table~\ref{table:EF}. The relative weight of the different bands is indicated by the inset bar chart. The weighted distribution of group velocity magnitudes over all the different bands is indicated by a black dashed line, and the mean value is indicated by a red dotted vertical line.
	}
	\label{fig:A1}
\end{figure}

\begin{figure}
	\centering
	\includegraphics[width=16cm]{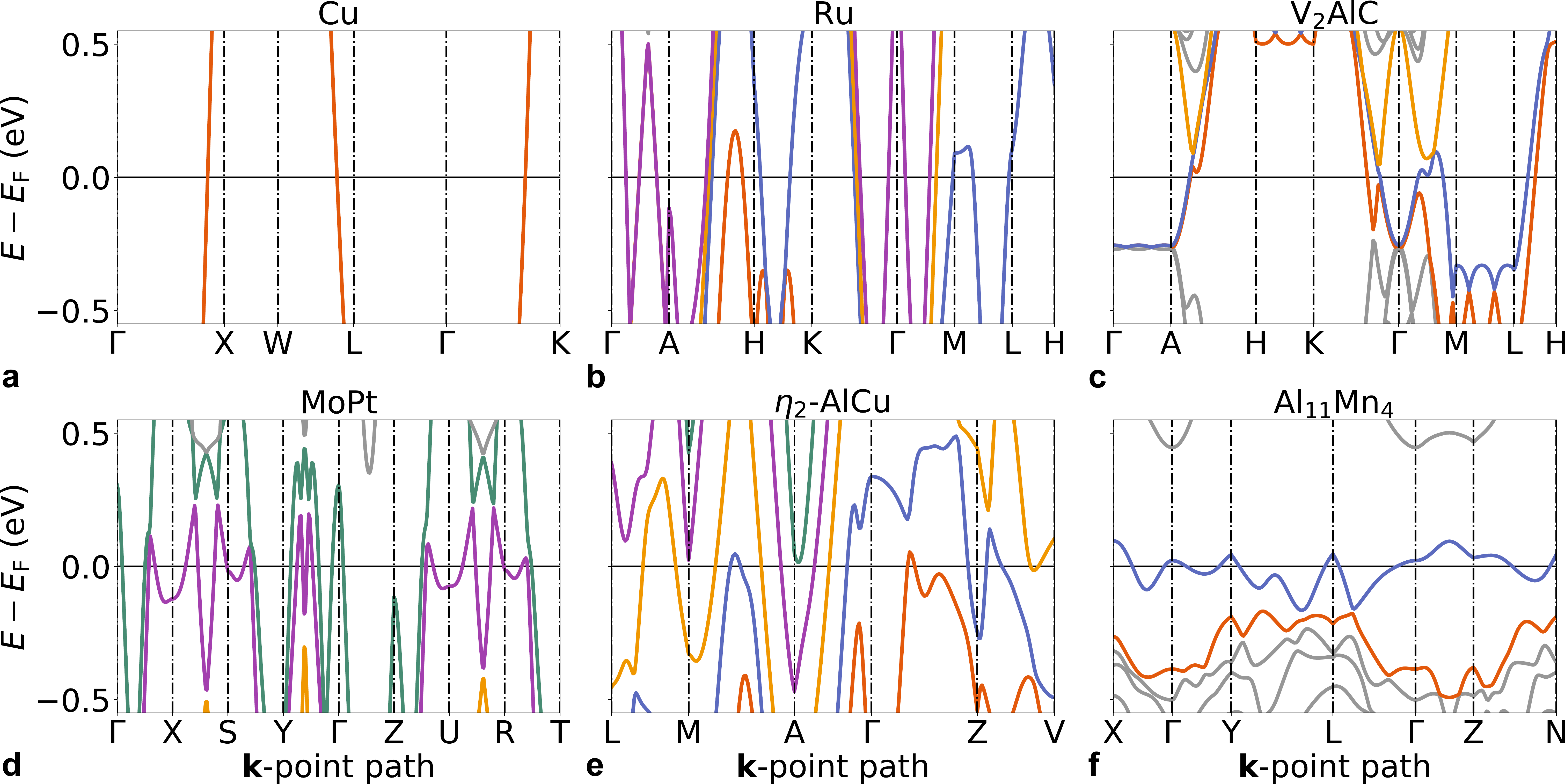}
	\caption{
		The electronic band structures of the example metals near the Fermi energy. The bands that are considered in the velocity distributions in Fig.~\ref{fig:A1} are indicated with the same color as the distribution profile.
	}
	\label{fig:A2}
\end{figure}

A detailed overview of the velocity magnitude distributions of the different conduction bands near the Fermi level is presented for the different example metals in Fig.~\ref{fig:A1} (see Fig.~\ref{fig:A2} for the corresponding band structures). The distribution within each band is presented, as well as the total weighted distribution for all (conduction) bands with solutions near the Fermi energy. The weighted distributions have been obtained with Fermi-Dirac statistics at $T = 300\,\textnormal{K}$.

\section{\label{sec:numerical_integration}Numerical integration}

For the numerical integration of Eqs.~\eqref{eq:rho_lambda} and \eqref{eq:rho_tau}, we start from a collection of $N_\text{I} \times N_\text{II} \times N_\text{III}$ $\veck$ points over the primitive cell in reciprocal space for each (conduction) band with index $n$, with the corresponding energies $E^{(n)}(\veck)$ obtained from density functional theory calculations (see Sec.~\ref{subsec:band_structure}):

\begin{equation}
	\veck = \mu_\text{I} \veck_\text{I} / N_\text{I} + \mu_\text{II} \veck_\text{II} / N_\text{II} + \mu_\text{III} \veck_\text{III} / N_\text{III} \qquad \mu_{\text{I},\text{II},\text{III}} \in \{ 0, 1, \ldots, N_{\text{I},\text{II},\text{III}} \},
\end{equation}

\noindent with $\veck_{\text{I}, \text{II}, \text{III}}$ the primitive vectors in reciprocal space (see Table~\ref{table:EF}).
We interpolate the energies between the coordinates $(\mu_\text{I}/N_\text{I}, \mu_\text{II}/N_\text{II}, \mu_\text{III}/N_\text{III})$ with third-degree polynomials, obtaining a continuous energy spectrum $E^{(n)}(k_{x'}, k_{y'}, k_{z'})$, with $k_{x',y',z'} \in [0, 1]$. As the primitive vectors are not orthogonal, these coordinates are not Cartesian. By performing the following linear transformation, we can write the interpolated energies in terms of the Cartesian coordinates $(k_x, k_y, k_z)$:

\begin{equation}
	(k_x', k_y', k_z') = (k_x, k_y, k_z) \cdot \begin{pmatrix} k_{\text{I},x} & k_{\text{I},y} & k_{\text{I},z} \\ k_{\text{II},x} & k_{\text{II},y} & k_{\text{II},z} \\ k_{\text{III},x} & k_{\text{III},y} & k_{\text{III},z} \end{pmatrix}^{-1}.
\end{equation}

Having obtained interpolated functions for the energies $E^{(n)} (\veck)$ over the full primitive cell in reciprocal space in this way, we can calculate the velocity $\vecv^{(n)}(\veck) = \hbar^{-1} \nabla_\veck E^{(n)}(\veck)$ and numerically integrate the tensors. For the integration over the Brillouin zone, we consider a Monte Carlo integration scheme with a high sampling density near the Fermi surface to account for the Fermi-Dirac weighting. This is achieved by randomly generating a large number of points $\veck^{(n)}_\nu$ ($\nu = 1, \ldots, N_\text{points}$) for each band with $E_\textsc{f} - \Delta E/2 \leq E^{(n)}(\veck^{(n)}_\nu) \leq E_\textsc{f} + \Delta E/2$, effectively occupying a total volume $\mathcal{V}^{(n)}(\Delta E)$ in reciprocal space (which is estimated by constructing a 3D mesh from the set of points). The integration of a Fermi-Dirac weighted average over the Brillouin zone for a particular band is then numerically implemented as:

\begin{equation}
	\int_\textsc{bz} \text{d}^3 \veck \, f^{(n)}(\veck) \left. \frac{\deriv f_\textsc{fd}(\epsilon)}{\deriv \epsilon} \right|_{\epsilon = E^{(n)}(\veck)} \approx \frac{\mathcal{V}^{(n)}(\Delta E)}{N_\text{points}} \sum_{\nu=1}^{N_\text{points}} f^{(n)}(\veck^{(n)}_\nu) \left. \frac{\deriv f_\textsc{fd}(\epsilon)}{\deriv \epsilon} \right|_{\epsilon = E^{(n)}(\veck^{(n)}_\nu)},
\end{equation}

\noindent here considering an arbitrary function $f^{(n)}(\veck)$. For the results presented here, we considered $N_\text{points} = 10^5$ and $\Delta E = 0.4$ eV, which was found to be sufficient for the results to converge.

\clearpage

\end{document}